\documentclass[10pt,twocolumn,letterpaper]{article}

\usepackage{cvpr}
\usepackage{times}
\usepackage{epsfig}
\usepackage{graphicx}
\usepackage{amsmath}
\usepackage{amssymb}
\usepackage{float}
\usepackage{multirow,array}
\usepackage{multirow}
\usepackage{hhline}
\usepackage{titling}

\usepackage[breaklinks=true,bookmarks=false]{hyperref}
\date{December 13, 2019}
\cvprfinalcopy 

\def\cvprPaperID{****} 

\newcommand*\samethanks[1][\value{footnote}]{\footnotemark[#1]}

\title{Reading Between the Demographic Lines: \\Resolving Sources of Bias in Toxicity Classifiers}
\author{
  Elizabeth Reichert\thanks{Equal Contribution, Stanford University}\\
  \texttt{ecreich@stanford.edu}
  \and
  Helen Qiu\samethanks\\
  \texttt{shiqiu21@stanford.edu}
  \and
  Jasmine Bayrooti\samethanks\\
  \texttt{jbayrooti@stanford.edu}
}
\begin{document}

\maketitle

\begin{abstract}
The censorship of toxic comments is often left to the judgment of imperfect models. Perspective API, a creation of Google technology incubator Jigsaw, is perhaps the most widely used toxicity classifier in industry; the model is employed by several online communities including \textit{The New York Times} to identify and filter out toxic comments with the goal of preserving online safety. Unfortunately, Google's model tends to unfairly assign higher toxicity scores to comments containing words referring to the identities of commonly targeted groups (e.g., ``woman,'' ``gay,'' etc.) because these identities are frequently referenced in a disrespectful manner in the training data. As a result, comments generated by marginalized groups referencing their identities are often mistakenly censored. It is important to be cognizant of this unintended bias and strive to mitigate its effects. To address this issue, we have constructed several toxicity classifiers with the intention of reducing unintended bias while maintaining strong classification performance. 

\end{abstract}

\section{Introduction}

\textbf{Summary of approach.}
In this paper, we describe a variety of toxic comment detection models crafted with the goal of lessening identity-driven bias while preserving high classification performance. To balance our dataset, we used over-sampling and under-sampling techniques in combination with natural text generation. We created logistic regression, neural network, and LSTM (Long Short-Term Memory) models to perform the toxicity classification of comments. We explored different features (including TF-IDF and GloVe embeddings), hyper-parameter configurations, and model architectures. We fed into our models the original training data as well as our re-balanced dataset and compared their performances both in terms of classification and mitigation of unintended bias. We also compared the performance of our best model to Perspective API by examining the prevalence of false positives in identity-related tweets by members of the U.S. House of Representatives that are assumed to be non-toxic.

\textbf{Social impact.} We hope that the techniques explored in this paper will help to improve the fairness of toxic classifiers. Mistakenly flagging non-toxic content, especially content produced by marginalized voices, is an important social issue; it is critical to foster a safe online environment while also protecting the freedom of expression of all people.

\section{Related Work}
Prior work in the realm of toxic speech detection (e.g. detecting offensive speech in tweets) has used bag-of-words approaches for feature representation along with models in logistic regression, decision trees, and linear SVMs to classify public social comments [1, 2]. More recent work has explored the application of CNN (Convolutional Neural Network) and GRU (Gated Recurrent Neural Network) in detecting hateful language on Twitter [3]. However, a common challenge with these models is the misclassification of content associated with frequently targeted communities [4]. To address Perspective API’s unintended bias problem, Jigsaw has proposed bias mitigation methods including mining assumed ``non-toxic'' data from Wikipedia articles to achieve a greater balance between toxic and non-toxic content referencing identity terms [5]. The optimal method to eliminate unintended bias resulting from imperfect training sets is not yet clear. According to a study on racial bias in abusive language detection, a one-size-fits-all approach to defining and detecting abusive language is not viable due to different speech norms in different communities, and providing more context, such as the author’s profile and the content that the author is responding to, is key to improving model performance [6].

\section{Data}

\subsection{Datasets}

\textbf{Toxicity-labeled comment data.} Our \underline{\href{https://www.kaggle.com/c/jigsaw-unintended-bias-in-toxicity-classification/data?fbclid=IwAR3_st3fwI7wJ4c2Bs1aHW6g6gattGq9wOPY3kACnPnbBjrN8NQi2R0UP8w}{dataset}} consists of approximately $1.8$ million comments from the currently-inactive online social platform Civil Comments. Each comment is labeled with a target value\footnote{The target is the fraction of human labelers that found the comment ``very toxic'' (meaning severely hateful or disrespectful to the point of making one very likely to leave the platform) or ``toxic'' (meaning unreasonable or rude to the point of making one somewhat likely to leave the platform).} between $0$ and $1$ (inclusive) as well as several additional toxicity sub-type attributes (i.e., obscene, threat, insult, severe toxicity, identity attack, sexually explicit). Approximately $400,000$ comments are also marked with identity attributes representing the identities mentioned in the comment (e.g., ``Muslim,'' ``bisexual,'' etc.). 

We assign a binary toxicity label to each comment in our dataset, where the label ``non-toxic'' ($0$) corresponds to a toxicity score of below $0.5$ and the label ``toxic'' ($1$) corresponds to a toxicity score of between $0.5$ and $1$. The following are examples of comments from our dataset including toxicity scores and labels:
\begin{itemize}
    \item ``This bitch is nuts. Who would read a book by a woman.''; Toxicity Score: $0.83$; Toxicity Label: $1$ (``toxic'')
    \item ``Why would you assume that the nurses in this story were women?''; Toxicity Score: $0.0$; Toxicity Label: $0$ (``non-toxic'')
\end{itemize}

To speed up training, we randomly sampled $50\%$ of the $1.8$ million comments in our dataset to construct the training and test sets based on an $80:20$ split.

\textbf{Identity-related tweets of politicians.} To measure the degree of identity-driven bias mitigation achieved by our top-performing model relative to Perspective API, we utilized a \underline{\href{https://www.kaggle.com/kapastor/democratvsrepublicantweets\#ExtractedTweets.csv}{dataset}} of tweets made by members of the U.S. House of Representatives. The dataset consists of the latest $200$ tweets (as of May 17th, 2018) from all members of the House of Representatives. A total of $84,502$ tweets are included along with the tweeter's name and political party affiliation\footnote{Approximately $51\%$ of tweets are made by Republicans while around $49\%$ of tweets are authored by Democrats.} (i.e., Democrat or Republican). After filtering for tweets referencing at least one identity, $1,984$ tweets remained. We used this set of approximately $2,000$ identity-related tweets as an additional dataset to compare our best model's level of unintended bias relative to that of Google's Perspective API model.

\subsection{Features}
\textbf{Term-frequency times inverse document-frequency (TF-IDF).} We use scikit-learn's TF-IDF vectorizer, which is a normalized bag-of-words transformation of sentences, to construct features. The resulting vector for each comment is of length $SIZE_{corpus}$ (i.e., $198,234$). Each element in the vector is the word frequency in the comment multiplied by the inverse of the word frequency in the dataset. The goal of using TF-IDF instead of the raw frequencies of a word occurrences is to lessen the impact of words that occur very often in a given corpus (e.g., stop words such as ``the,'' ``a,'' ``and,'' etc.) that have empirically less informative than words that occur less frequently in the corpus. For the features generated via TF-IDF, our original training input was of shape ($721950, 198234$) and the shape of our test set was ($180487, 198234$).

\textbf{GloVe embeddings.} We experimented with pre-trained GloVe embeddings as another way to extract features from the training data, since prior work suggests that GLoVe outperforms related models on similarity tasks and named entity recognition [7]. We used the $25$-dimension GloVe vectors trained from $2$ billion tweets on Twitter. We chose to use this specific GloVe embeddings model because the twitter data that the model was trained on is similar to the comment data on Civil Comments. For the features generated via GloVe embeddings, our original training input was of shape ($721950, 25$) and the shape of our test set was ($180487, 25$).

\section{Data Exploration}
It is important to note that our dataset is imbalanced. Using a toxicity score threshold of $0.5$, there are roughly $8\%$ toxic examples and $92\%$ non-toxic examples.

Out of the approximately $1.8$ million comments in our dataset, $405,130$ of those comments were randomly selected and marked with identity attributes representing the identities mentioned in the comment (e.g., ``black,'' ``Christian,'' etc.). We found that, out of the comments that were labeled with identity attributes, approximately $44\%$ did not reference any identities while around $56\%$ referenced at least one identity—roughly an even split. 

We found that comments that referenced at least one identity were more likely to be toxic than comments that did not reference any identities. The distribution of comments based on their toxicity and reference of identity terms is given in Figure 1.\footnote{``Toxic'' refers to comments with a toxicity score of greater than or equal to $0.5$ while ``non-toxic'' refers to comments with a toxicity score less than $0.5$. ``Identity'' refers to comments that reference at least one identity term and ``non-identity'' refers to comments that do not reference any identity terms. An identity term is referenced if the fraction of human labelers that believe an identity term is referenced is greater than $0$.}

\begin{figure}[H]
    \centering
    \includegraphics[width=0.35\textwidth]{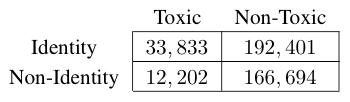}
    \caption{Distribution of comment categories.}
\end{figure}

We notice that there is a higher incidence of toxicity in comments that reference identity terms than comments that do not reference any identity terms. Specifically, we observe that toxic comments are $14.9\%$ of comments that reference identities and only $6.8\%$ of comments that do not reference identities. This difference is highly statistically significant with $p < 0.00001$. Since we observe such a large difference in the prevalence of toxicity in identity versus non-identity comments, it is important that, when re-sampling comments to balance the split of toxic versus non-toxic examples in our dataset, we control for the split of identity versus non-identity examples within the toxic and non-toxic examples. To do so, we sample our data equally from each the following comment categories: toxic identity, toxic non-identity, non-toxic identity, and non-toxic non-identity.

\section{Methods}
\subsection{Baseline vs. Oracle}
Our baseline is a Naive Bayes Classifier (NBC) whose input is a bag-of-words representation of a pre-processed comment. Due to limitations in computing power, we randomly sampled $25\%$ of our dataset (approximately $450,000$ comments) to construct our training and test sets. Performing an $80:20$ random split of the data, we trained with $360,976$ comments and ran the NBC on a test set of size $90,244$. We performed data pre-processing and feature extraction in scikit-learn. We observe that the Naive Bayes binary classification model performs poorly, with precision of approximately $15\%$ and recall of around $5\%$. This means that out of all of the toxic comments, our model only predicts $5\%$ of them as being toxic, and out of all of the comments our model predicts as toxic, only $15\%$ of them are actually toxic. The F1 score of this model is roughly $7.5\%$.

The oracle is the human labeling of the toxicity of the comments. Humans are better able to comprehend the intended meaning of the comment and the contextual use of identity terms. The bag-of-words approach de-contextualizes comments by eliminating their sequential relationships, and the Naive Bayes model is not sophisticated enough to capture the meaning of the text—rather it attempts to discern the underlying distribution that produced the model’s inputs under the assumption that the inputs are independent.

\subsection{Logistic Regression}
Logistic regression is commonly applied to classification problems and is a simple, yet powerful model. We implemented two logistic regression models: one with features generated via TF-IDF and one with features generated via GloVe embeddings. The output for the model was binary—either $0$ (non-toxic) or $1$ (toxic).

\subsection{Neural Network}
Neural networks have strong representational power; neural networks with at least one hidden layer are “universal approximators” and can approximate any continuous function [9]. We first trained a two-layer fully-connected neural network that has the fully-connected—ReLU—fully-connected architecture. The score for each output class is calculated with $f = \max(0, x W_1 + bias))W_2 + bias$. Here, the input $x$ is a $N \times D$ matrix where $N$ represents the number of training examples in the training set, $D$ represents the number of features per training example, and $W_1$ is a $D \times SIZE_{hidden}$ matrix that transforms the training data into a $N \times SIZE_{hidden}$ intermediate matrix. The ReLU activation function $\max(0, xW_1)$ is a non-linearity that is applied element-wise.  Finally, $W_2$ is a matrix of size $SIZE_{hidden}\times C$, and thus $f$ has shape $N \times C$. The model outputs $C$ numbers interpreted as class scores for each training example. The parameters $W_1$ and $W_2$ are learned with stochastic gradient descent where the gradients are derived with chain rule and computed with back propagation.

We used scikit-learn to create the neural network architecture. We first trained the two-layer neural net on scikit-learn's TFIDF-vectorized features. The output layer produced classification scores for two classes—non-toxic represented by $0$, and toxic represented by $1$. The size of the hidden layer was $100$ and the learning rate was $1 \times e^{-5}$. We also experimented with transforming the comments into GloVe embeddings, summing up the GloVe vectors for each words in a comment and feeding in the sum vector as input. Once again, the output for the model was binary—either $0$ (non-toxic) or $1$ (toxic).

\begin{figure}[H]
    \centering
    \includegraphics[width=0.35\textwidth]{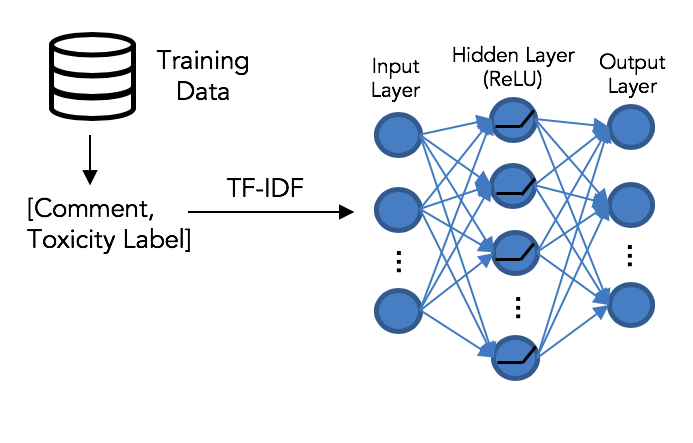}
    \caption{Illustration of two-layer neural network architecture.}
\end{figure}
\break
To explore if a slightly deeper neural architecture would result in better model performance, we also trained a three-layer fully-connected neural network, with hidden layer sizes of $75$ and $50$ and a learning rate of $1 \times e^{-5}$. The input and output were of the same shape as the two-layer model.

\subsection{Two-Layer Bidirectional LSTM}
Words that appear near the beginning of a sentence often influence the meaning of subsequent words, ultimately affecting the interpretation of that sentence. Therefore, to distinguish between appropriate versus inappropriate usage of identity terms, it is critical to process the context in which these identity words are used in each comment. We implemented a two-layer bidirectional LSTM (Long Short-Term Memory) model to capture word dependencies and sequential information. We chose a bidirectional architecture since it encapsulates information in both the forward and backward sequences of words in sentences. 

\begin{figure}[H]
    \includegraphics[width=0.5\textwidth]{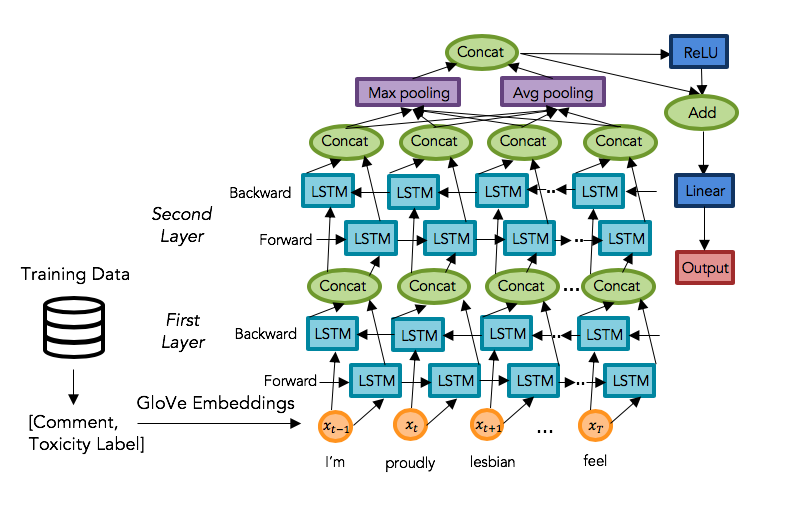}
    \caption{Illustration of two-layer bidirectional LSTM architecture.}
\end{figure}

The model takes in a padded list of GloVe vectors as input, where each vector is a $25$-dimensional GloVe representation of a word in the comment. We dropped or padded each comment so that all comments had a length of $200$. The input to the first-layer LSTM has shape $(batch$ $size$, $comment$ $length$, $embed$ $length)$. The output of the bidirectional processing by the first-layer LSTM $lstm1$ is concatenated and fed into the second-layer LSTM $lstm2$. The output from $lstm2$ is then concatenated and put through a max pooling layer and an average pooling layer. We used both max pooling and average pooling and concatenated their results with the goal of extracting the most extreme word information as well as preserving the overall semantic information of each comment. Finally, with two linear layers we reduce the dimension of the output associated with each comment to $1$, which is then converted via sigmoid function to a value between $0$ to $1$ representing the probability of the comment being toxic.

To prevent overfitting, we experimented with spatial dropout regularization on the input to $lstm1$. Using hyperparameter tuning, we selected $128$ as the number of LSTM hidden units, $512$ as the batch size, $1 \times e^{-5}$ as the learning rate, and $0.3$ as the dropout rate. After each epoch in training, we utilized our validation set of $40,000$ comments to evaluate the model's F1 score and update our best model, which was later used for prediction on the test set. Our implementation of LSTM drew inspiration from \underline{\href{https://www.kaggle.com/bminixhofer/simple-lstm-pytorch-version}{this code}}. 

To determine the appropriate number of epochs, we plotted training loss against epoch value. For both the LSTM model trained on the original training set and the LSTM model trained on the balanced training set, after approximately 10 epochs, we observed that training loss converged, thus suggesting that a local optimum had been reached.

\begin{figure}[H]
    \centering
    \includegraphics[width=0.4\textwidth]{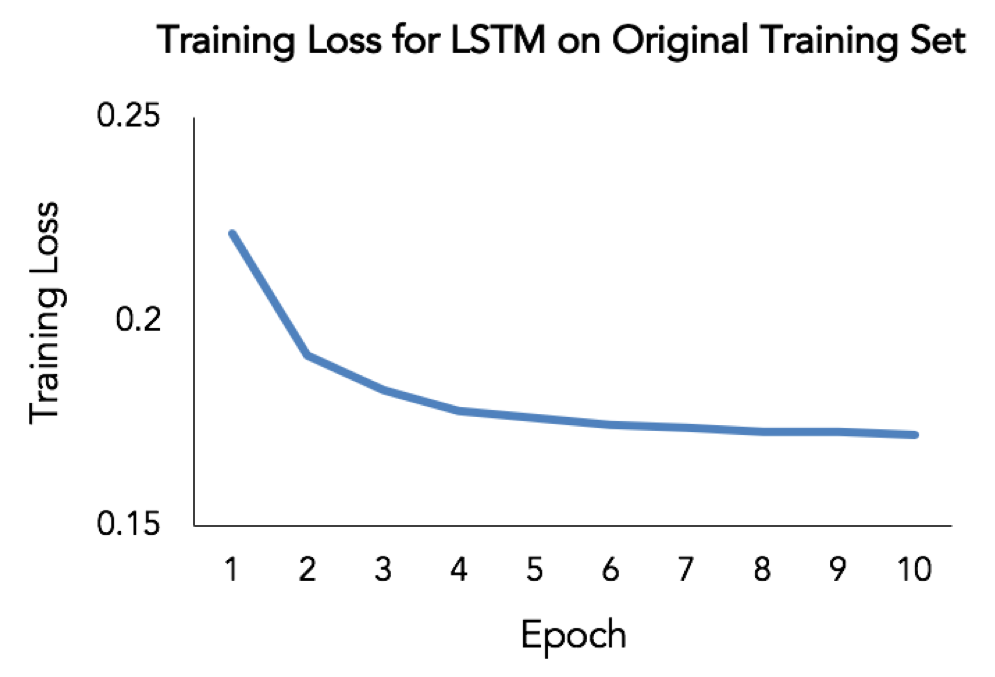}
    \caption{Training loss for LSTM converges to $0.17$ for original training set.}
\end{figure}

\begin{figure}[H]
    \centering
    \includegraphics[width=0.4\textwidth]{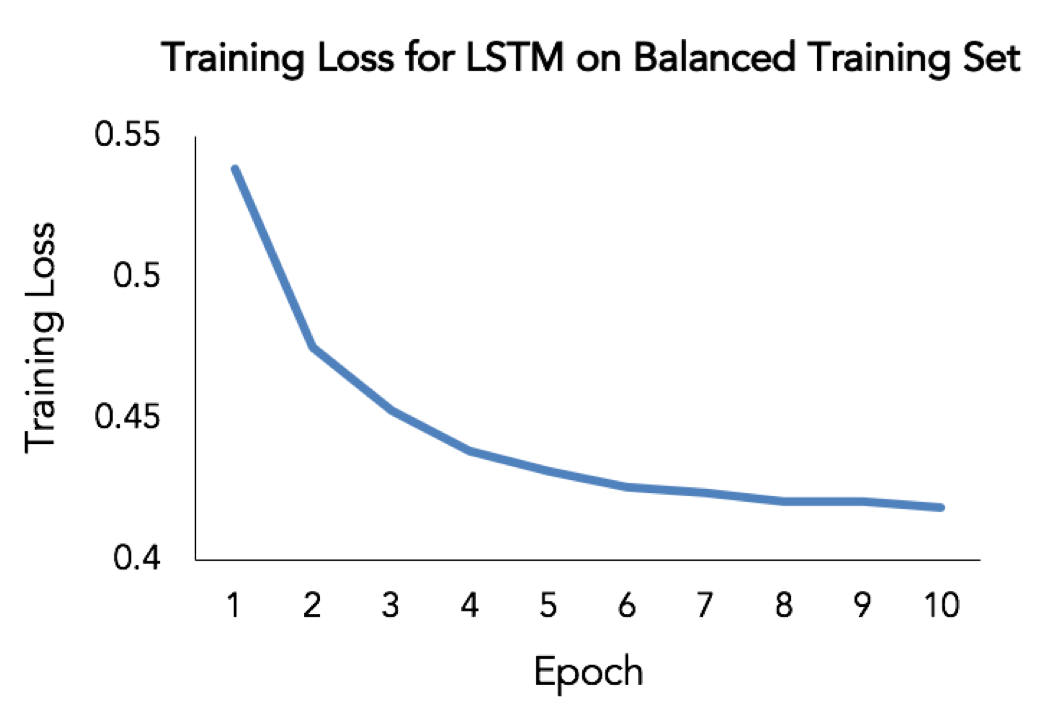}
    \caption{Training loss for LSTM converges to $0.42$ for balanced training set.}
\end{figure}

\subsection{Data Rebalancing}
We used natural text generating techniques to produce additional comments to aid in balancing our training dataset. More specifically, we created our own comment templates for the following comment categories: toxic non-identity, toxic identity, non-toxic identity.\footnote{Non-toxic non-identity comments covered too broad of a topic range for their generation to be worthwhile.} We then filled in these templates with different identity terms (as well as identity slurs when the comments were designed to be offensive). Using these comments as source text, we utilized a bidirectional LSTM encoder and LSTM decoder to generate variations (or paraphrases) of these comments with temperature 0.8 [11]. The following are comments generated by the paraphraser:
\begin{itemize}
    \item I am black and I feel that this issue is important to me.
    \item I'm black and I think this problem is significant to me.
\end{itemize}
We then incorporated these comments into our training set and used over-sampling and under-sampling techniques to sample $50,000$ comments from each of the comment categories (i.e. toxic identity, toxic non-identity, non-toxic identity, and non-toxic non-identity). Thus, our rebalanced training set consisted of a total of $200,000$ comments. We ran the logistic regression, neural network, and LSTM models mentioned above again on the rebalanced training set, and tested the model performances on the original test set so that we were able to directly compare the performance of the models trained on both the original and rebalanced training sets.

\section{Results}

\subsection{Evaluation Metrics} 

\textbf{Classification performance.} We used area under the receiver operating characteristic curve (AUC) and F1 score\footnote{F1 score is defined as the harmonic mean of precision and recall.} on the test set as measures of model performance. Both metrics are commonly used to evaluate model performance on imbalanced data sets.

\textbf{Mitigation of identity-driven unintended bias.} To measure the bias of our classifier, we compared the false positive rate ($FPR$) for comments that reference identities to that of comments that do not reference identities. We also compared the false negative rate ($FNR$) for identity versus non-identity comments. By the Equality of Odds, these rates should not differ if the model is fair [11].

\textbf{Comparison to Perspective API.} To evaluate the ability of our model to mitigate identity-driven unintended bias compared to Google's toxicity classifier Perspective API, we examined tweets of U.S. Representatives referencing identities (e.g., ``transgender,'' ``Jewish''). We assume that politicians would not tweet toxic content, and thus we used the prevalence of false positives in this dataset as a measurement of unintended bias.

\subsection{Model Performance}

\textbf{Models trained on original training set.} We begin by evaluating the performance of our models trained on the original (unbalanced) training set.

\begin{figure}[H]
    \includegraphics[width=0.5\textwidth]{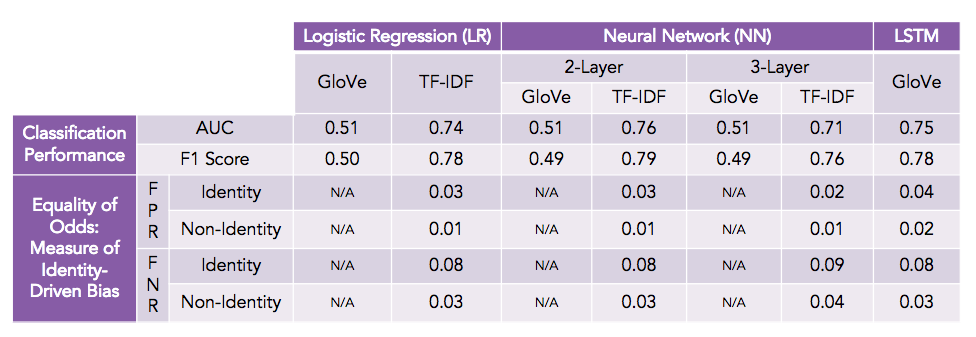}
    \caption{Overview of the performance of our models trained on the original (imbalanced) training set and tested on the test set.}
\end{figure}

Before rebalancing, the two-layer neural network with TF-IDF features achieved superior classification performance with an AUC of $0.76$ and an F1 score of $0.79$. For all of our models, compared to TF-IDF vectorization, feature extraction using GLoVe embeddings resulted in reduced classification performance. This is likely because summing up the GLoVe-represented word vectors in a sentence leads to a loss of semantic contributions from individual words. GLoVe representations were more suitable for our LSTM model to learn word dependencies and sequential relationships, as evidenced by strong model performance; the two-layer bidirectional LSTM achieved an AUC of $0.75$ and an F1 score of $0.78$. Surprisingly, our three-layer neural network performed worse than our two-layer neural network. To determine if our three-layer neural network was overfit, we compared both the AUC and F1 score on the training set versus test set. We found that the AUCs and F1 scores are essentially equivalent for both datasets, suggesting that the worse performance is not due to overfitting. It is important to note that our logistic regression model, the simplest of all of our models, achieved comparable performance to our best model, with an AUC of $0.74$ and an F1 score of $0.78$, when both models used TF-IDF features.

For each of our models, we notice that the false positive rate ($FPR$) for comments that reference identity terms is $2$ to $3$ times greater than the $FPR$ for comments that do not reference identity terms. Similarly, we see that, for each of our models, the false negative rate ($FNR$) for ``identity'' comments is $2$ to $3$ times greater than the $FNR$ for ``non-identity'' comments. Thus, since the $FPR$ and $FNR$ differ between the two groups (``identity'' and ``non-identity'' comments), by the definition of fairness outlined in the Equality of Odds, we observe bias [12].

\textbf{Models trained on rebalanced training set.} Next, we evaluate the performance of our models trained on the rebalanced training set.

\begin{figure}[H]
    \includegraphics[width=0.5\textwidth]{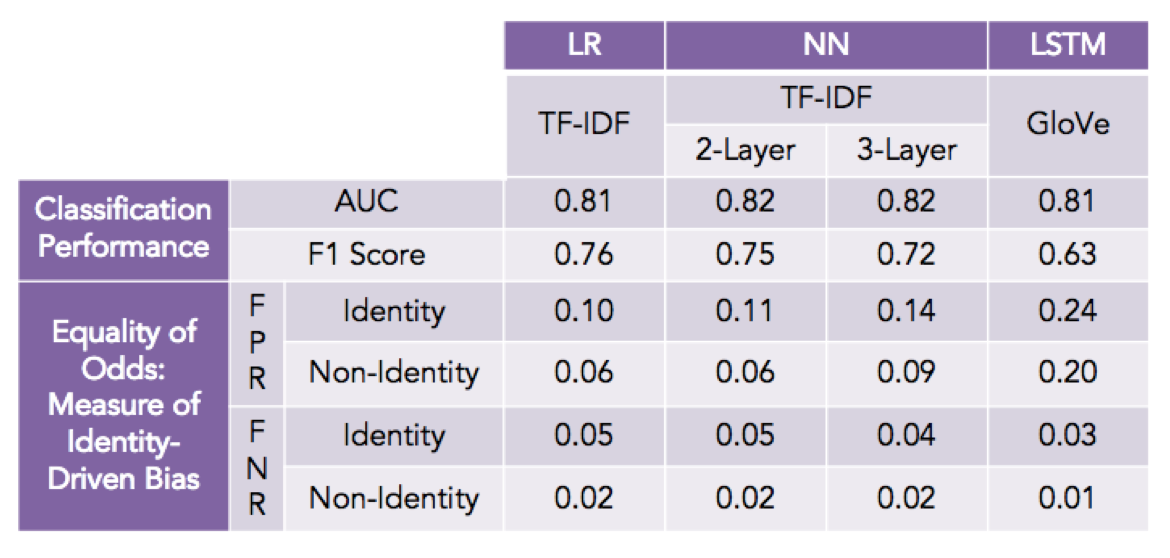}
    \caption{Overview of the performance of our models trained on the rebalanced training set and tested on the test set.}
\end{figure}

After rebalancing, the two-layer neural network with TF-IDF features achieved the highest AUC ($0.82$), while the logistic regression model trained on TF-IDF features gave the highest F1 Score ($0.76$). Since GloVe embeddings did not perform well for our logistic regression and neural network models, we decided that when training on the rebalanced dataset, we would not use features via GloVe embeddings for those models. All of our models were similar in terms of classification performance, with AUCs ranging from $0.81$ to $0.82$. The F1 scores of our models are also very much alike (ranging from $0.72$ to $0.76$) with the exception of the LSTM model. Our lower F1 score for this model may be the result of a sub-optimal classification threshold\footnote{The toxicity classification threshold is $0.5$.} since F1 score is dependent on the binary classification cutoff while AUC is not. We believe that the LSTM model could have outperformed our other models if it was supplied more training data. Since LSTM models have more hyperparameters, training on a larger dataset may have resulted in better tuning of those parameters.

It makes sense that the $FPR$s are higher for our models trained on the rebalanced dataset than the $FPR$s for our models trained on the original training set, since, in the process of rebalancing, we are adding more toxic examples. We notice that, when comparing the $FPR$ and $FNR$ of the ``identity'' comments versus ``non-identity'' comments for each of our models, the ratio of the two values has decreased (instead of $2$ to $3$ times greater, the ratios of the rates are $1$ to $2.5$ times greater), suggesting unintended bias has been reduced. It is important to note that the increase in $FPR$s is larger in magnitude than the decrease in $FNR$s for our models trained on the balanced training set; ideally, rebalancing would have resulted in a smaller ratio between $FPR$s and $FNR$s for identity and non-identity comments in addition to reducing the magnitude of these rates.

Our models trained on the rebalanced dataset perform better in terms of AUC on the test set than our models trained on the original training set. This suggests that rebalancing training data is an effective way to improve model classification performance. When comparing the top-performing models in terms of AUC, we observe an increase from $0.76$ to $0.82$.

\begin{figure}[H]
    \centering
    \includegraphics[width=0.5\textwidth]{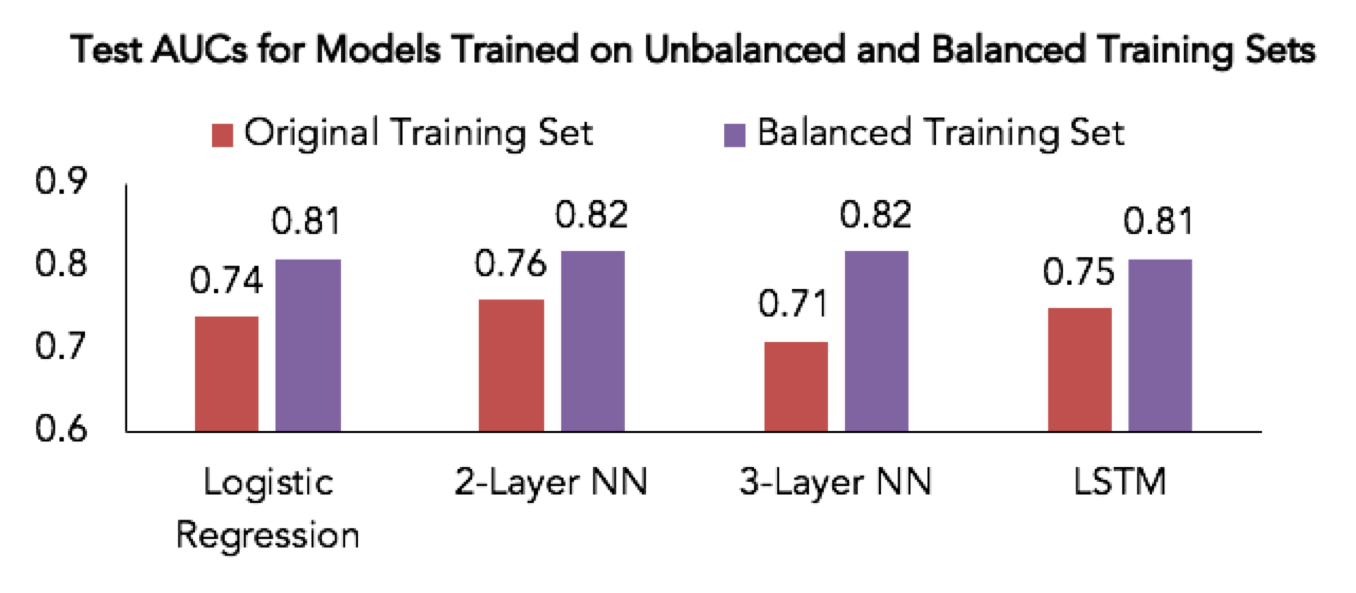}
    \caption{Balancing training set increases test AUC for all models.}
\end{figure}

\textbf{Effects of incremental data rebalancing.} We examined the effects of incrementally rebalancing the training set on the classification performance and mitigation of unintended bias on our top-performing model (i.e., the two-layer neural network). We incrementally rebalanced each of the four comment categories (i.e., non-toxic identity, non-toxic non-identity, toxic identity, and toxic non-identity) in batches of $10,000$ comments until the dataset was fully balanced.\footnote{The dataset is fully balanced when each of the comment categories has $50,000$ comments. The number of non-toxic comments was incrementally decreased and the number of toxic comments was incrementally increased to reflect the directionality of the rebalancing we performed on our original (imbalanced) dataset.}

We observe that, for all rebalancings, the AUC on the test set increases as the dataset becomes more balanced. When adding more toxic examples, unsurprisingly, we notice that $FPR$ increases while $FNR$ decreases; this is because increasing the number of toxic examples in our training set makes the model more likely to predict examples in the test set as toxic. Using the definition of fairness given by Equality of Odds, we visualize the mitigation of bias as the convergence of the $FPR$s and the convergence of the $FNR$s for identity and non-identity comments [11].

\begin{figure}[H]
    \centering
    \includegraphics[width=0.2\textwidth]{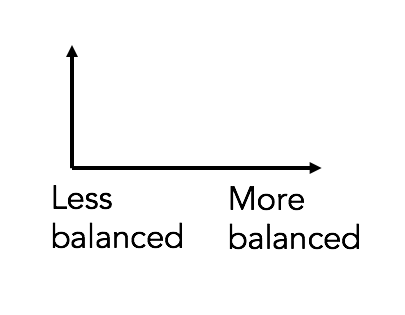}
    \caption{Key for figures $10$ through $13$; moving rightwards along the $x$-axis increases the degree of balance in the training set.}
\end{figure}

When rebalancing the number of non-toxic identity comment examples, we observe an increase in the test AUC with F1 score remaining constant. We notice that the $FPR$s slightly diverge while the $FNR$s slightly converge.

\begin{figure}[H]
    \includegraphics[width=0.5\textwidth]{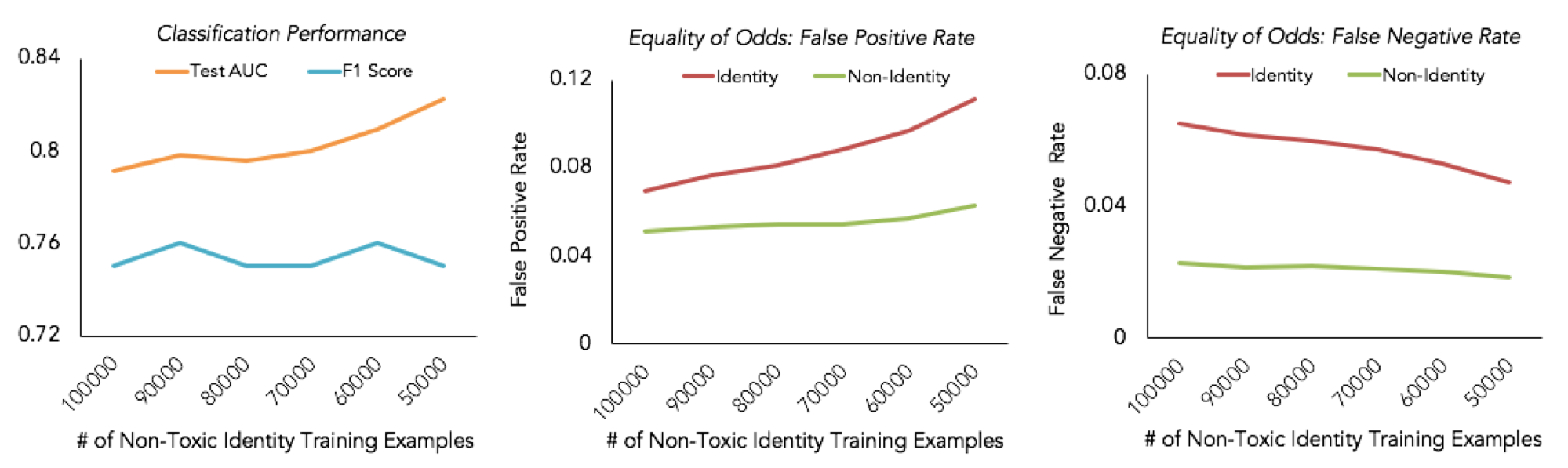}
    \caption{Incrementally rebalancing the number of non-toxic identity comments.}
\end{figure}

When rebalancing the number of non-toxic non-identity comment examples, we observe an increase in the test AUC and F1 score. We see that the $FPR$s marginally converge while the $FNR$s marginally diverge.

\begin{figure}[H]
    \includegraphics[width=0.5\textwidth]{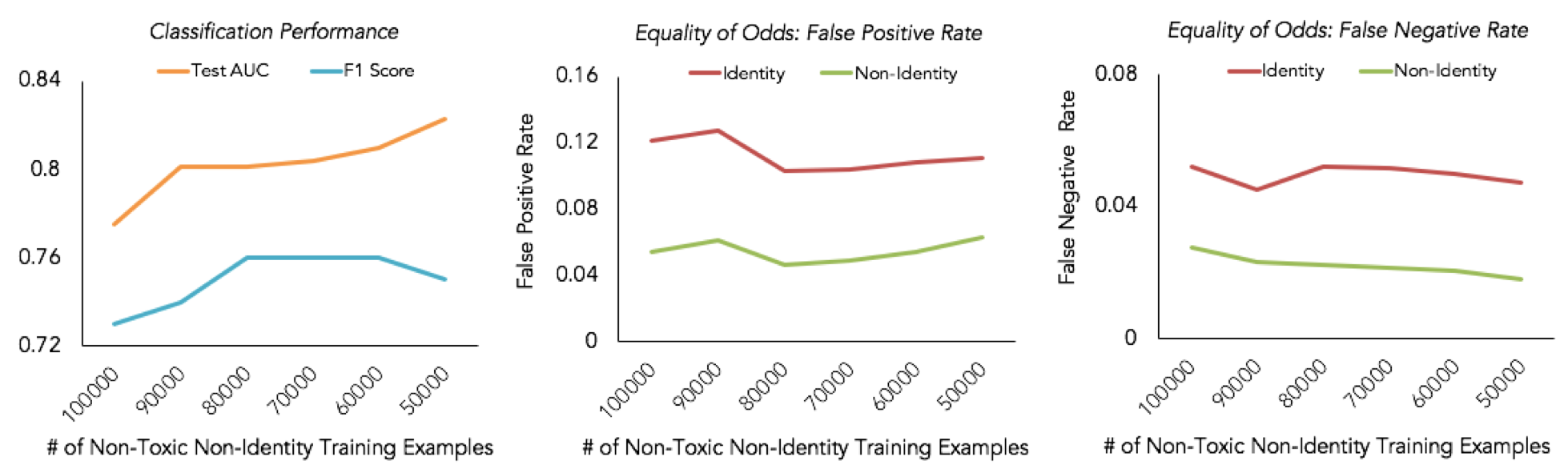}
    \caption{Incrementally rebalancing the number of non-toxic non-identity comments.}
\end{figure}

We find that the two-layer neural network performs worst in terms of AUC when there are no toxic identity comment examples. This makes sense since toxic comments commonly reference identity terms in a threatening or disrespectful manner. The test AUC increases dramatically as we rebalance the number of toxic identity comments while the F1 score slightly increases. The $FPR$s converge, intersecting at $10,000$ toxic identity comments, and then diverge while the $FNR$s substantially converge. As expected, the $FPR$s increase and the $FNR$s decrease.

\begin{figure}[H]
    \includegraphics[width=0.5\textwidth]{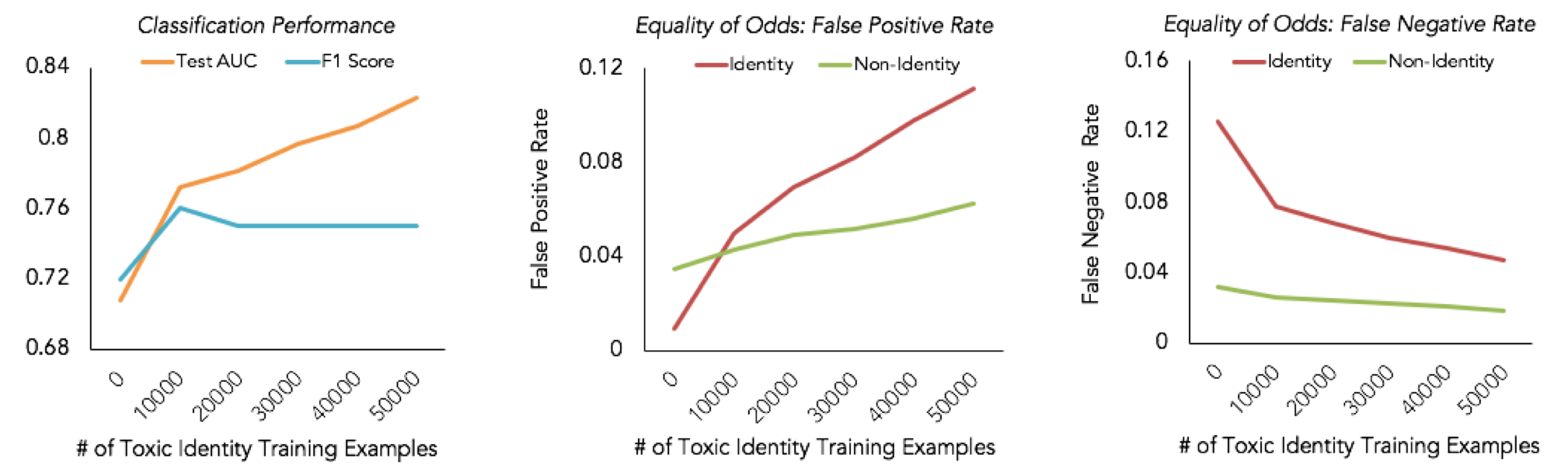}
    \caption{Incrementally rebalancing the number of toxic identity comments.}
\end{figure}

The test AUC increases while the F1 score remains constant when rebalancing the number of toxic non-identity comments. The $FPR$s converge while increasing, and the $FNR$s essentially remain equidistant while decreasing.

\begin{figure}[H]
    \includegraphics[width=0.5\textwidth]{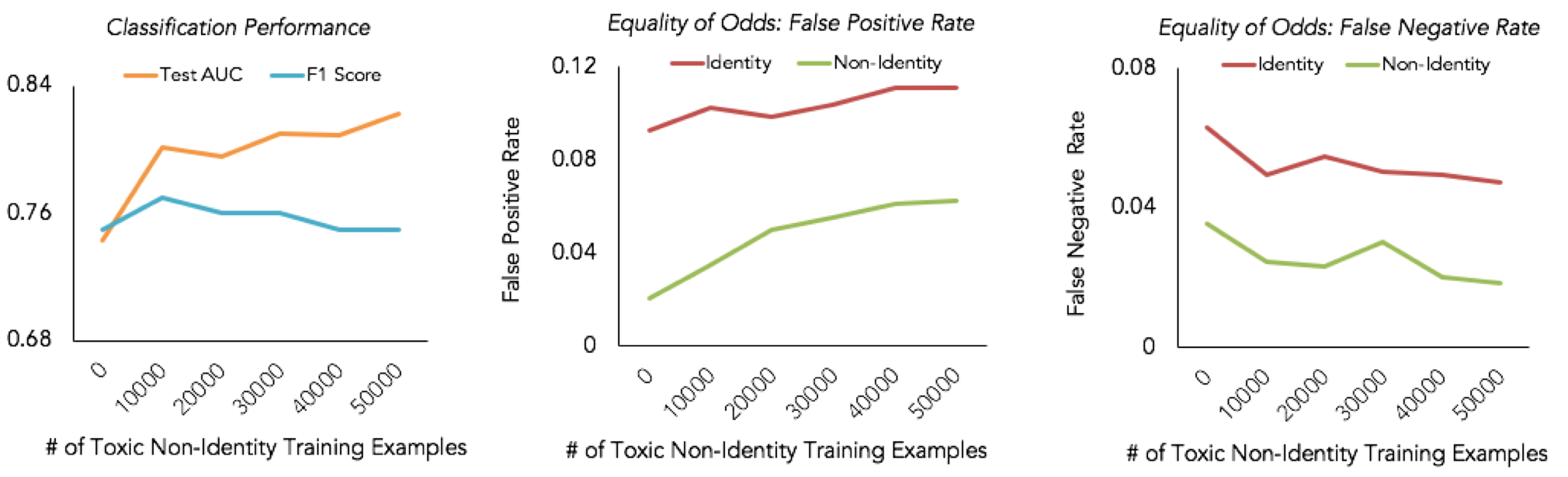}
    \caption{Incrementally rebalancing the number of toxic non-identity comments.}
\end{figure}

\textbf{Bias mitigation comparison to Perspective API.} In the dataset of $1,984$ tweets made by U.S. politicians referencing identities that are assumed to be non-toxic, Perspective API labeled $10.3\%$ as toxic, while our best model (i.e., the two-layer neural network) only labeled $6.9\%$ as toxic. The lower prevalence of false positives achieved by our model indicates that we have outperformed Perspective API in terms of reducing identity-driven bias. We do not claim that our model performs generally better in terms of classification performance than Perspective API (it is doubtful since Google's toxicity classifier has been trained on many more examples); however, this result does suggest that controlling for the reference of identity terms when rebalancing the training set to include an equal number of toxic and non-toxic examples is an effective means to mitigate identity-driven bias.

Out of the $1,984$ tweets in our test set, $145$ were incorrectly marked as toxic by Perspective API but correctly predicted to be non-toxic by our top-performing model. These tweets included the following:

\begin{itemize}
    \item We must hold hearings \& finally address that \#terrorism inflicted by white supremacy extremists is destroying \#USA.
    \item When bias drives discipline, black girls miss out on the chance to learn.
    \item \#WeRemember the victims of the Holocaust, including the millions of Jewish men, women, and children who were massacred.
\end{itemize}

After examining these tweets, it appears that Perspective API had difficulty distinguishing toxicity from negative sentiment in tweets that referenced identity terms. Tweets referencing identity terms commonly describe inequities between or acts of hate against identity groups. Although these tweets may be calling out unfairness or violence (topics that have negative sentiment), the content of these messages is not toxic.

Our model miscategorized $78$ tweets that Google's model correctly predicted. Selected examples are given below:

\begin{itemize}
    \item Today we recognize Transgender Day of Visibility, seeing transgender people for who they are where they are.
    \item To Jewish friends and family in NJ and around the world, I wish you all a Happy \#YomHaatzmaut.
    \item To all my Muslim brothers and sisters and those observing in the \#CA13, I wish you a blessed, peaceful and happy month of Ramadan.
\end{itemize}

The majority of tweets that our top-performing model incorrectly categorized had positive sentiment but referenced identity terms that were less common in our training data. Most of the identities in our training set related to gender (specifically women) and race, not sexuality and religion. We believe that having fewer examples of these identities hurt our model's ability to correctly identify these tweets as non-toxic. To achieve better performance, it would be beneficial to train our model on a more comprehensive set of identity comment examples. Although when generating identity comment examples we included a wide range of identity terms, since our text generation was limited to paraphrasings, we were not able to achieve a breadth of positive contexts in which these identity terms are referenced.

\section{Future Work}
We hope to explore models and techniques that will allow us to further reduce unintended bias while improving classification performance. We would like to explore more sophisticated models such as BERT in combination with LSTM, which incorporates self-attention mechanisms to make connections between words and relevant context and identify dependencies of words that are far away from each other [13]. We also wish to investigate more complex text generation techniques to increase the size and variety of our training set examples since having a greater volume of training examples will be useful when training more complex models.

\section{Acknowledgements}
We would like to acknowledge Hancheng Cao for his superb mentorship throughout the research process; we thank him for his expert advice and encouragement. We would also like to recognize Victor Suthichai, whose paraphraser we customized to generate additional comment examples when balancing our training set and Benjamin Minixhofer, whose code we referenced when designing the architecture of our LSTM model.

\section{References}
[1] Chatzakou, Despoina, et al. ``Mean birds: Detecting aggression and bullying on twitter.'' Proceedings of the 2017 ACM on Web Science Conference. ACM, 2017.

[2] Davidson, Thomas et al. ``Automated Hate Speech Detection and the Problem of Offensive Language.'' ICWSM (2017).

[3] Zhang, Ziqi et al. ``Detecting Hate Speech on Twitter Using a Convolution-GRU Based Deep Neural Network.'' ESWC (2018). 

[4] Burnap, Peter and Matthew Leighton Williams. ``Cyber Hate Speech on Twitter: An Application of Machine Classification and Statistical Modeling for Policy and Decision Making.'' (2015).

[5] Dixon, Lucas et al. ``Measuring and Mitigating Unintended Bias in Text Classification.'' AIES (2018).

[6] Davidson, Thomas et al. ``Racial Bias in Hate Speech and Abusive Language Detection Datasets.'' ArXiv abs/1905.12516 (2019).

[7] Pennington, Jeffrey et al. ``Glove: Global Vectors for Word Representation.'' EMNLP (2014).

[8] Cybenko, G. ``Approximation by Superpositions of a Sigmoidal Function.'' Math. Control Signals Systems (1989)2: 303-314.

[9] Sap, Maarten et al. ``The Risk of Racial Bias in Hate Speech Detection.'' ACL (2019).

[10] Minixhofer, Ben. ``Simple LSTM - PyTorch.'' Kaggle, 2019. https://www.kaggle.com/bminixhofer/simple-lstm-pytorch-version

[11] Suthichai, Victor. ``Paraphraser.'' Github, 2017. https://github.com/vsuthichai/paraphraser

[12] Moritz Hardt, Eric Price, and Nathan Srebro. 2016. ``Equality of Opportunity in
Supervised Learning.'' CoRR abs/1610.02413 (2016). http://arxiv.org/abs/1610.
02413

[13] Jacob Devlin Ming-Wei Chang Kenton Lee Kristina Toutanova. 2019. ``BERT: Pre-training of Deep Bidirectional Transformers for Language Understanding'' arXiv:1810.04805v2 [cs.CL] 24 May 2019

\end{document}